%% file: main.tex
\documentclass[twocolumn]{aastex6}

\usepackage{natbib}
\usepackage{hyperref}

\newcommand\mJyperbm{{$\rm mJy\,beam^{-1}$}}

\newcommand\um{{$\mu \rm m$}}
\newcommand\kms{{$\rm km s^{-1}$}}
\newcommand\vlsr{{v_{\rm LSR}}}
\newcommand\planck{{\it Planck}}

\newcommand\idl{{\sc IDL}}
\newcommand\starlink{{\sc Starlink}}
\newcommand\cupid{{\sc cupid}}
\newcommand\clumpfind{{\sc ClumpFind}}

\newcommand\fmean{{F_\mathrm{m}}}
\newcommand\fe{{F_\mathrm{e}}}
\newcommand\nc{{n_\mathrm{c}}}
\newcommand\sd{{\sigma_\mathrm{std,meas}}}

\newcommand\sdfid{{\sigma_\mathrm{std,fid}}}
\newcommand\ucal{{u_\mathrm{cal}}}
\newcommand\noise{{\sigma_{\mathrm{rms}}}}
\newcommand\Lsun{{L_{\odot}}}
\newcommand\Msun{{M_{\odot}}}

\fullcollaborationName{...}

\begin{document}


\title{Submillimeter continuum variability in Planck Galactic cold clumps}



\author{Geumsook Park\altaffilmark{1}}
\author{Kee-Tae Kim\altaffilmark{1}}
\author{Doug Johnstone\altaffilmark{2,3}}
\author{Sung-ju Kang\altaffilmark{1}}
\author{Tie Liu\altaffilmark{1,4}}
\author{Steve Mairs\altaffilmark{4}}
\author{Minho Choi\altaffilmark{1}}
\author{Jeong-Eun Lee\altaffilmark{5}}
\author{Patricio Sanhueza\altaffilmark{6}}
\author{Mika Juvela\altaffilmark{7}}
\author{Miju Kang\altaffilmark{1}}
\author{David Eden\altaffilmark{8}}
\author{Archana Soam\altaffilmark{9}}
\author{Julien Montillaud\altaffilmark{10}} 
\author{Gary A. Fuller\altaffilmark{11}} 
\author{Patrick M. Koch\altaffilmark{12}} 
\author{Chang Won Lee\altaffilmark{1,13}}
\author{Dimitris Stamatellos\altaffilmark{14}} 
\author{Jonathan Rawlings\altaffilmark{15}}
\author{Gwanjeong Kim\altaffilmark{16}} 
\author{Chuan-Peng Zhang\altaffilmark{17}} 
\author{Woojin Kwon\altaffilmark{1,13}}
\author{Hyunju Yoo\altaffilmark{1}}


\altaffiltext{1}{Korea Astronomy and Space Science Institute, 
776 Daedeokdae-ro, Yuseong-gu, Daejeon 34055, Republic of Korea; pgs@kasi.re.kr}

\altaffiltext{2}{NRC Herzberg Astronomy and Astrophysics, 5071 West Saanich Rd, 
Victoria, BC, V9E 2E7, Canada}

\altaffiltext{3}{Department of Physics and Astronomy, University of Victoria, 
Victoria, BC, V8P 5C2, Canada}

\altaffiltext{4}{East Asian Observatory, 660 North A'ohoku Place, 
University Park, Hilo, Hawaii 96720, USA}

\altaffiltext{5}{School of Space Research, Kyung Hee University, 1732, Deogyeong-Daero, Giheung-gu Yongin-shi, Gyunggi-do 17104, Republic of Korea}

\altaffiltext{6}{National Astronomical Observatory of Japan, National Institutes of Natural Sciences, 2-21-1 Osawa, Mitaka, Tokyo 181-8588, Japan}

\altaffiltext{7}{Department of Physics, P.O. Box 64, FI-00014, University of Helsinki, Finland}

\altaffiltext{8}{Astrophysics Research Institute, Liverpool John Moores University, IC2, Liverpool Science Park, 146 Brownlow Hill, Liverpool L3 5RF, UK}

\altaffiltext{9}{SOFIA Science Centre, USRA, NASA Ames Research Centre, MS-12, N232, Moffett Field, CA 94035, USA}

\altaffiltext{10}{Institut UTINAM, CNRS UMR 6213, OSU THETA, Universit\'e Bourgogne Franche-Comt\'e, France}

\altaffiltext{11}{UK ALMA Regional Centre Node, Jodrell Bank Centre for Astrophysics, School of Physics and Astronomy, The University of Manchester, Oxford Road, Manchester M13 9PL, UK}

\altaffiltext{12}{Institute of Astronomy and Astrophysics, Academia Sinica. 11F of Astronomy-Mathematics Building, AS/NTU No. 1, Section 4, Roosevelt Rd., Taipei 10617, Taiwan}

\altaffiltext{13}{University of Science and Technology, Korea (UST), 217 Gajeong-ro, Yuseong-gu, Daejeon 34113, Republic of Korea}

\altaffiltext{14}{Jeremiah Horrocks Institute for Mathematics, Physics \& Astronomy, University of Central Lancashire, Preston PR1 2HE, UK}

\altaffiltext{15}{Department of Physics and Astronomy, University College London, Gower Street, London, WC1E 6BT, UK}

\altaffiltext{16}{Nobeyama Radio Observatory, National Astronomical Observatory of Japan, National Institutes of Natural Sciences, Nobeyama, Minamimaki, Minamisaku, Nagano 384-1305, Japan}

\altaffiltext{17}{National Astronomical Observatories, Chinese Academy of Sciences, Beijing, 100012, Peopleʼs Republic of China}


\begin{abstract}

In the early stages of star formation,
a protostar is deeply embedded 
in an optically thick envelope such that it is not directly observable.
Variations in the protostellar accretion rate, however, will cause luminosity changes 
that are reprocessed by the surrounding envelope and are observable at submillimeter wavelengths.
We searched for submillimeter flux variability 
toward 12 Planck Galactic Cold Clumps detected by the James Clerk Maxwell
Telescope (JCMT)-SCUBA-2 Continuum Observations of Pre-protostellar Evolution (SCOPE) survey. 
These observations were conducted at 850~\um\ 
using the JCMT/SCUBA-2. 
Each field was observed three times over about 14 months
between 2016 April and 2017 June.
We applied a relative flux calibration
and achieved a calibration uncertainty of $\sim 3.6$\% on average.
We identified 136 clumps across 12 fields and detected four sources with flux variations of $\sim 30\%$.
For three of these sources, the variations appear to be primarily due to large-scale contamination,
leaving one plausible candidate.
The flux change of the candidate may be associated with low- or intermediate-mass star formation assuming a distance of 1.5~kpc,
although we cannot completely rule out the possibility that it is a random deviation.
Further studies with dedicated monitoring would provide a better understanding of 
the detailed relationship between submillimeter flux and accretion rate variabilities
while enhancing the search for variability in star-forming clumps farther away than the Gould Belt.

\end{abstract}





\section{Introduction} \label{sec:intro}

A protostar gains mass by accreting material 
through a protostellar disk embedded in a circumstellar envelope \citep[for a review, see e.g.][]{hartmann2016}. 
Understanding the mass accretion of protostars is an essential component 
to characterizing their overall formation and evolution.
The earliest formulations of star formation theory assumed 
a steady-state accretion model where the amount of mass gained 
by the protostar was constant over time \citep{shu1977, terebey1984}.
Many subsequent observational studies \citep[e.g.,][]{kenyon1990,evans2009}, however,
indicate that protostellar luminosities are lower than 
predicted by these conventional models.
One solution to this so-called ``luminosity problem'' 
is a variable protostellar accretion rate (often called episodic accretion), 
where bright outbursts occur over short timescales 
and the forming star spends most of its time in a
``quiescent'' phase \citep{dunham2010, dunham2012}. 
The evolutionary lifetime of protostars, however, is still uncertain 
and refining the current estimates may also contribute to correcting
this apparent discrepancy between the protostellar luminosity predicted by models 
and the current observations \citep[e.g.,][]{evans2009,mckee2011}.
Ultimately,
studies of the variability of accretion rates are critical 
in order to understand the physics of the circumstellar disk
and how the mass is transported onto the protostar, itself.

In this study, we focus on detecting signs of episodic accretion 
in the earliest stages of star formation. 
The majority of accretion variability observations have so far been carried out
in the evolved stages of pre-main-sequence stars \citep[e.g.,][]{kospal2007,aspin2009,  
caratti2011, covey2011, fischer2012, reipurth2012}. 
EX Lupi \citep[e.g.,][]{herbig2008, aspin2010} 
and FU Orionis \citep[e.g.,][]{herbig1977, hartmann1996} 
sources could be classical examples of episodic accretion 
occurring after the deeply embedded phase. 
The spectacular observational change in optical brightness for these objects is 
about a factor of 10 or more and lasts for several months to decades.
Recently, however, a few outbursts from deeply embedded protostellar objects 
have been reported \citep[e.g.,][]{safron2015, hunter2017, yoo2017}.

The mass accretion rates are high 
in the early stages of protostellar evolution \citep[e.g.,][]{whitworth2001,schmeja2004},
so  we expect the accretion variability to be more significant than during later stages. 
Direct observations at optical or near-infrared (near-IR) variability of protostellar systems (star(s)/disk(s)) are, however, 
very challenging because these systems are heavily embedded in optically thick, dense envelopes.
Thus, indirect observations at submillimeter wavelengths are now being explored.
\citet{johnstone2013} analyzed the flux variability of a protostellar envelope caused by 
 outbursts of the central source 
using far-IR and submillimeter continuum emission.
The model suggests that mid- to far-IR observations would be ideal 
to detect variability changes over timescales of hours to days.
The study also revealed that detecting variability at submillimeter wavelengths
should be achievable, although variations occur 
over longer timescales of approximately one month.
There are, indeed, some examples of known flux variations
that were detected in submillimeter continuum emission:
about a factor of 2 flux increase at 350~\um\ and 450~\um\
toward the Class~0 source HOPS~383 \citep{safron2015}.
More recently, \citet{mairs2018} reported
that HOPS~358 now shows a strong, declining light curve over the course of 16 months.
The Class~I protostar, EC~53, displayed a 50\% flux increase at 850~\um\ \citep{yoo2017}.
Furthermore,
\citet{hunter2017} reported  
high-mass protostellar system NGC~6334I-MM1 
with a factor of 4.2 increase in 870~\um\ continuum interferometric flux
and a 30\% increase in the submillimeter single-dish flux.

As one of the large programs at the East Asian Observatory's James Clerk Maxwell Telescope (JCMT),
the JCMT Transient Survey \citep{herczeg2017} has been designed 
to search for this type of long-term variability in submillimeter dust emission
surrounding deeply embedded protostars in eight nearby star-forming regions within $\sim 500$~pc from the Sun.
The Transient Survey is the only monitoring survey performed at submillimeter wavelengths. 
The first major results were released after 1.5~yr and 
the team has also used archival data to identify variability 
over a timescale of $\sim5$~yr \citep{johnstone2018, mairs2017a}.
They found that $\sim10\%$ of deeply embedded protostars display 
varying flux at the level of 5--10\% per year. 
However, these nearby regions are mostly forming low-mass stars.

The SCUBA-2 Continuum Observations of Pre-protostellar Evolution (SCOPE)
survey with the JCMT \citep{liu2018a} 
has observed $\sim1200$ \planck\ Galactic cold clumps \citep[PGCCs;][]{planck2016}, 
which were selected in wide ranges of Galactic longitudes and latitudes.
The SCOPE sample was biased to high column density PGCCs of 
$N_{\mathrm H_2} > 1 \times 10^{21}\,{\rm cm}^{-2}$ (in \planck\ measurements),
but also included randomly selected lower column density PGCCs 
($> 5 \times 10^{20}\,{\rm cm}^{-2}$ in \planck\ measurements).
For about 3/5 of the SCOPE sample, 
physical properties are given in \citet{planck2016}:
(1) about 70\% among them are concentrated within 1~kpc 
while the others are widely distributed at up to $\sim 8$~kpc,
with an average angular size of $\sim 8$\arcmin;
(2) the mass range is from $0.1\,\Msun$ to $10^5\,\Msun$
(see Figure~2 of \citealt{liu2018a} and Figure~1 of \citealt{eden2019} for detailed distributions).
Therefore, the SCOPE sample contains diverse clumps in different Galactic environments,
from low-mass to high-mass star-forming regions at various distances from the Sun.
In addition,
to obtain deep images of high-mass star-forming regions
as well as to detect large flux variation events,
the SCOPE survey observed some ($< 30$) PGCCs, 
which are composed of multiple substructures,
on three separate occasions.
We note that the SCOPE survey looks at more distant clumps than the JCMT Transient Survey and,
thus, they are more likely to contain groups of protostars rather than individuals.
When accretion variability is detected, therefore,
the flux can be diluted if the event originates in a single protostar 
as the beam contains many protostars.
Nevertheless, as shown in previous studies by \citet{mairs2017b} and \citet{johnstone2018},
we expect to uncover flux variability at about the 10\%-level or larger.

In this paper, we examine the flux variability from 
12 PGCC fields in the first quadrant of the Galactic plane
using the SCOPE survey data that are described in Section~\ref{sec:obs}.
The data reduction, including calibration and clump identification, is 
presented in Section~\ref{sec:reduc}.
We follow the procedures of
the Transient Survey team \citep[e.g.,][]{mairs2017b, johnstone2018} 
with appropriate modifications.
We present the results of the examination of flux variability 
toward identified clumps in Section~\ref{sec:res} 
and discuss possible candidates of flux variation in Section~\ref{sec:dis}.
Section~\ref{sec:sum} summarizes the main results.

\section{Data} \label{sec:obs}

The SCOPE survey mapped approximately 1200 PGCCs  \citep{planck2016} at 850~\um\ 
using SCUBA-2, the submillimeter continuum imaging instrument \citep{holland2013} at the 15m JCMT.
The survey was begun in 2015 December and was completed in 2017 July.
Each map is about 12\arcmin\ in diameter,
and the main beam size (FWHM) of JCMT/SCUBA-2 is 
14\farcs1 at 850~\um\ \citep{dempsey2013}.
Each field was observed 
under grade 3/4 weather conditions with zenith opacities at 225~GHz between 0.1 and 0.15.
The first released data were obtained by filtering out scales larger than 200\arcsec\
in order to remove the effects of the atmosphere, which is bright in the submillimeter regime. 
The pixel size is 4\arcsec.
The applied flux conversion factor (FCF) is 
554~$\rm Jy\,pW^{-1}\,beam^{-1}$ \citep{liu2018a},
which is slightly higher than
the usual FCF of 537~$\rm Jy\,pW^{-1}\,beam^{-1}$ given by \citet{dempsey2013}.
Considering this research is a part of the SCOPE survey, 
we initially adopted the FCF values calculated from \citet{liu2018a} 
in order to keep the consistency of the data.
Nevertheless,
since we performed a relative flux calibration (see Section~\ref{sec:reduc} for details),
the absolute flux calibration is not essential for this study.
This survey provides $\sim 20$~times higher angular resolution images compared to
the \planck\ 353~GHz ($\sim 850$~\um) data ($\sim5\arcmin$).
Thus, 
complex substructures in the PGCCs can be resolved in the JCMT images 
which could not be resolved by \planck\ (e.g., see Figure~\ref{fig:coadd}).
\citet{liu2018a} presents a detailed description of the survey,
and \citet{eden2019} provides information of the first data release
and the catalog of compact sources resolved with the JCMT.

In this study, we selected 12 PGCC fields 
in the first quadrant of the Galactic plane that 
are moderately bright and contain a relatively large number of clumps.\footnote{
The definition of ``clump" is ambiguous.
In this paper, SCOPE clumps resolved in the JCMT images 
are at various distances (see Table~\ref{tab:fieldinfo})
and can contain substructures that are visible at higher resolution.
The SCOPE clumps shown in this paper 
encompass masses from tens of solar masses to thousands of solar masses,
spanning the range of cores to clouds.
For simplicity, we refer to all these objects as clumps.}
PGCCs are written using the acronym ``PGCCs'' in the text.
These regions span the Galactic longitude range of $14\arcdeg < l < 36\arcdeg$
and are located at heliocentric distances from $\sim 1.5$ to 17~kpc
(Table~\ref{tab:fieldinfo}).
The three observations of each field were not carried out with a regular cadence and, therefore,  
had intervals spanning three weeks to 13 months.
The total exposure time to complete each epoch is 15.4 minutes on average,
and the median and maximum of exposure times per pixel are 
$\sim 55$ and $\sim 200$~s, respectively. 
Each image was smoothed with a Gaussian kernel of 8\arcsec\ FWHM (twice the pixel size)
to reduce pixel-to-pixel noise.
Thus, the final images shown in this paper have an angular resolution of 
16\farcs2 FWHM  after smoothing.

\section{Data Reduction} \label{sec:reduc}

The default 850~\um\ absolute flux calibration produced by
the data reduction pipeline at the JCMT
yields a 5--10\% uncertainty in pointlike calibrator sources
over weather bands 1 through 4 \citep{dempsey2013, mairs2017b}. 
Therefore, to detect a 3$\noise$ change in the peak flux of a source, 
the brightness variation would need to be at least 15--30\%.
Simulations \citep[e.g.][]{bae2014, vorobyov2015} as well as 
JCMT Transient Survey observations \citep{mairs2017a, johnstone2018},
however, suggest that less dramatic flux variations are more common. 
In order to increase detection reliability, 
it is advantageous to calibrate the flux 
in a relative sense using the method presented by \citet{mairs2017b}. 
In this way, it is possible to reduce the
(relative) flux uncertainty to 2--3\%, 
which allows for statistically significant measurements of 
$\sim$6--10\% flux changes.

In our implementation of the relative flux calibration scheme, 
we restricted the sources 
with high ($> 25$) signal-to-noise ratios (S/Ns; see Section~\ref{sec:res-s1} for details).
However, unlike the Transient Survey procedure,
we did not require that the sources are compact.
In comparing the source fluxes of different epochs,
we checked that the locations of most peaks remain within the
nominal 2\arcsec--6\arcsec\ uncertainty of the JCMT pointing; 
a typical difference is $\sim 4\arcsec$.
Further, we applied an image registration technique.
We used the IDL/SUBREG procedure\footnote{\url{http://www.stsci.edu/~mperrin/software/sources/subreg.pro}} and 
derived offsets ($\Delta_\mathrm{ra}$ and $\Delta_\mathrm{dec}$) between different images.
The two-dimensional offsets ($\Delta = \sqrt{(\Delta_\mathrm{ra})^2+(\Delta_\mathrm{dec})^2}$) 
of our SCOPE fields have a mean of 4\arcsec\ and a standard deviation of 2\arcsec.
This is consistent with our previous, manual inspection.
For these reasons, in this research, the different peak positions in a given
clump area within the beam size, were assumed to originate from the same source.

As with \citet{mairs2017b},
the peak flux values were compared in order to determine the relative calibration.
For point sources, calibrating to the peak flux allows a single number (the relative flux calibration)
to be determined for each epoch,
independent of the many underlying physical aspects responsible for the original calibration uncertainty.
Sources of calibration uncertainty include: a poor measurement of the sky opacity,
changed throughput of the instrument, or a slight focus offset,
the latter of which will contribute to a change in the observed beam shape.
For extended sources, determining the relative calibration using only the peak flux introduces
an additional level of uncertainty since changes to the underlying beam profile
also produce changes in the expected flux of the source.
Despite this complication, \citet{mairs2017b} and \citet{mairs2015} found 
that the peak flux of bright sources embedded in extended emission are well-recovered and 
consistent for data reduction methods similar to those used in this study. 
Furthermore, we derived a robust uncertainty associated with the relative flux calibration factor (RFCF; calculated below)
as an additional check on the validity of the process.

\begin{figure*}[ht!]
\centering
\epsscale{1.}
\plotone{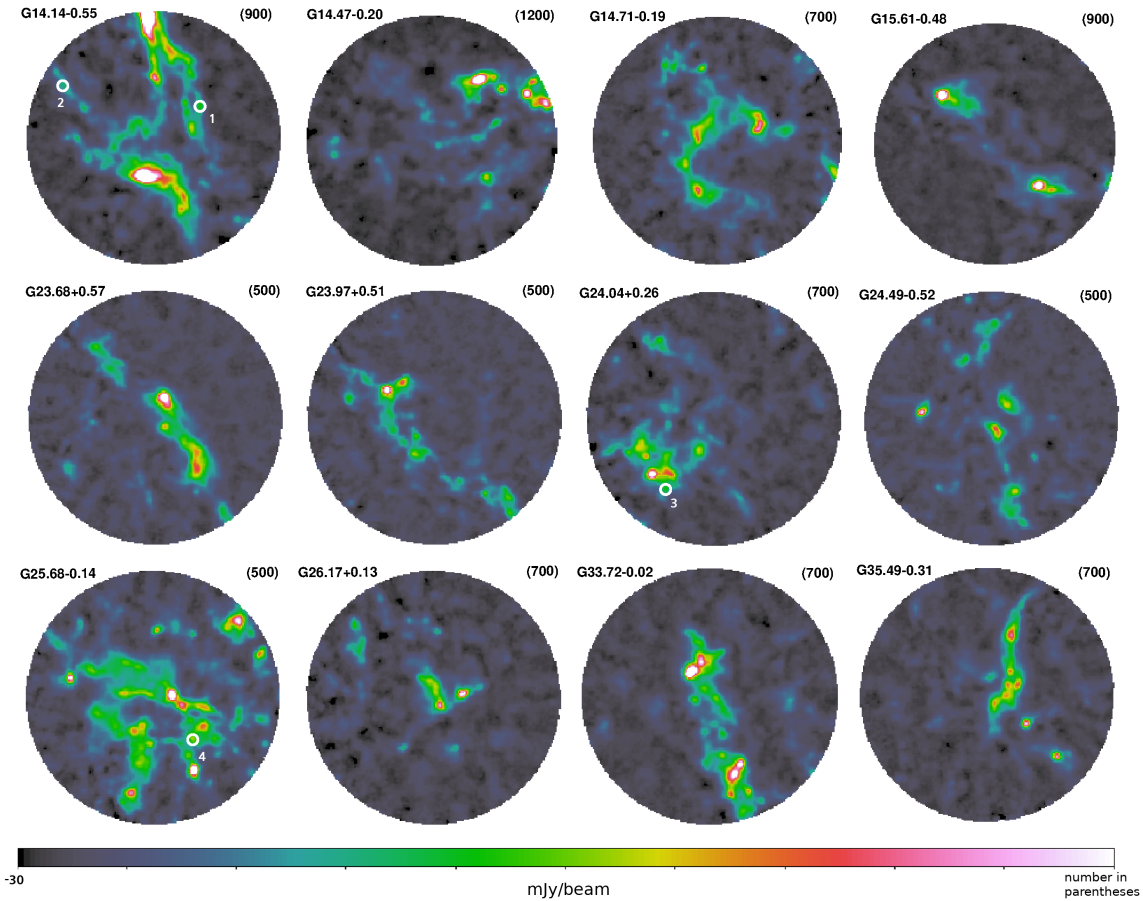}
\caption{
Co-added images for the 12 fields observed by the SCOPE survey.
Each image is cropped using a circle with a radius of 370\arcsec.
The field name is displayed at the top left of each image.
The color bar is shown in a linear scale, 
ranging from $-$30~\mJyperbm\ (black) to the value in parentheses
at the top right of each image.
White circles are marked to help to locate the outliers described in Section~\ref{sec:res}
and the number assigned to each circle is from Table~\ref{tab:out}.
\label{fig:coadd}}
\end{figure*}

In every observed field, each epoch was calibrated individually and co-added
to produce a deep, averaged image (Figure~\ref{fig:coadd}).
To achieve this, the {\sc{picard}} package \citep{gibb2013} found in 
the {\sc{Starlink}} software \citep{currie2014} was used. 
Although each co-added image was made by combining three epoch images,
the individual images were not very accurately aligned.
Therefore, we used the co-added image only
for the clump identification without getting into the details of alignment.
The peak flux density values used in the following analysis 
were obtained from the individual epochs.

We identified submillimeter clumps in the co-added images 
with the \clumpfind\ algorithm \citep{williams1994},
provided by {\sc{Starlink}}'s \cupid\ package \citep{berry2007},
considering an RMS noise level described in Section~\ref{sec:reduc-s1}.
There are several parameters to be set, 
such as ``FwhmBeam", ``MinPix", ``MaxBad", and ``Tlow."\footnote{
`FwhmBeam' defines the FWHM size of the JCMT beam in pixels,
which corresponds to 4.05 for our final images.
`MinPix' is the smallest number of pixels which a clump can have;
 we used a value of 13 as that corresponds to the area of a circle with a diameter equal 
to the (post-smoothing) beam FWHM.
`MaxBad' is the maximum fraction of blank pixels 
that can be contained in a clump, which is set to zero.
`Tlow' defines the lowest contour level to consider; we use $3\times$RMS noise.
A detailed description of the parameters is given at
\url{http://starlink.eao.hawaii.edu/docs/sun255.htx/sun255ss5.html}.} 
During the implementation, resultant clumps
containing fewer pixels than the area corresponding to the
beam size ($<$~MinPix) were discarded.
In addition, we excluded any clump if its peak is located beyond 370\arcsec\
from the central position of each map.\footnote{
This corresponds to the radius of the images in Figure~\ref{fig:coadd}.
The maps are shaped like an uneven circle
of which the radius extends from $\sim6\farcm5$ to $\sim8\arcmin$.
Near the edges of the images, the fields were much less exposed
and the coverage is uneven from epoch to epoch.
}
Information regarding the structure in each field, along with the derived 
relative calibration factors, are listed in Table~\ref{tab:fieldinfo}.

The relative flux calibration using the SCOPE data
started by assuming that none of the clumps are variable.
We found stable calibrator sources by an iterative method.
From the relative flux calibration derived using the stable sources,
we achieved a sensitivity that is sufficient to robustly detect a 10\% flux variation
(see Section~\ref{sec:reduc-s2} for details).
Then, we examined whether non-calibrator sources are outliers
and tested their significance with respect to the observational 
uncertainty.

\subsection{Measuring the Flux: Step 1} \label{sec:reduc-s1}

\begin{figure*}[ht!]
\epsscale{.7}
\plotone{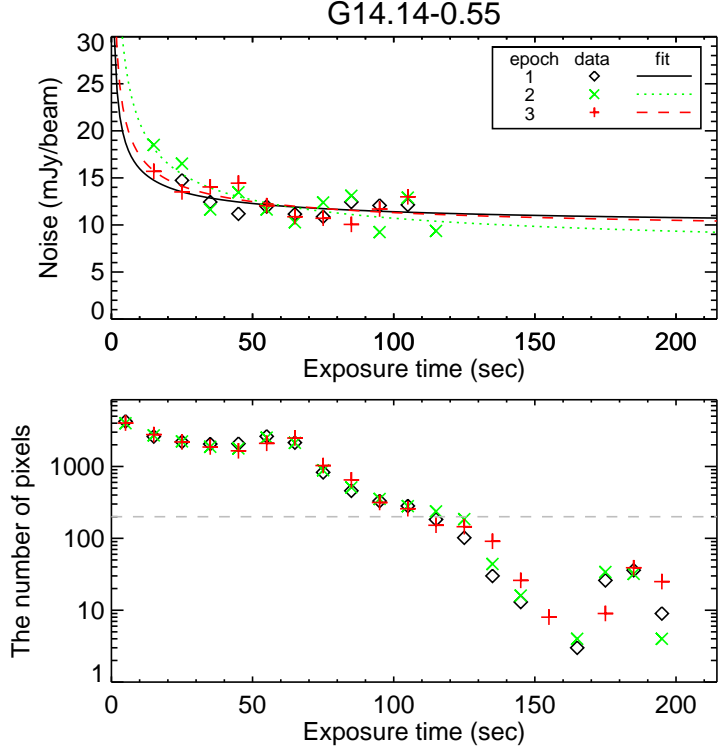}
\caption{
Top: example of the RMS noise as a function of exposure time.
Black diamonds, green crosses, and red pluses represent (in order) the 
three epochs observed of the G14.14$-$0.55 field. 
For the RMS noise calculation,
we only used bins where there are more than 200 pixels
(dashed gray line in the bottom panel).
The best-fit noise profiles $\sim 1/\sqrt{t}$,
where $t$ is the exposure time, are marked.
Their colors match the data points.
Bottom: the number of pixels as a function of exposure time.
\label{fig:exptime}}
\end{figure*}

A robust RMS noise measurement is important 
not only for identifying clumps 
but for assessing the significance of their flux variability. 
However, the RMS noise level of a CV daisy (CV = constant velocity) observation,\footnote{\url{http://www.eaobservatory.org/jcmt/instrumentation/continuum/ scuba-2/observing-modes/}}  
(the mode we employed in the SCOPE survey) is not uniform over the entire field.
Since SCUBA-2 generates a map of the exposure time for each mapping field,
we were able to use this map to characterize the RMS noise 
at different positions in the field. 
We measured the RMS noise level as a function of the exposure time 
in areas with no astronomical signal using data from each epoch.
The RMS noise levels showed gradual changes (almost flat) 
at exposure times larger than $\sim50$~s and 
increase sharply at shorter exposure times (see Figure~\ref{fig:exptime}).
By design, most pixels in the latter case are located near the edge of the uncropped images, 
so the data points with exposure times shorter than $\sim50$ seconds are insignificant
for our analysis.
We generated a best-fit noise profile for each epoch  
(curves in the top panel of Figure~\ref{fig:exptime})
using a simplified equation of the expected noise level ($y$)
$y = c_{1} + c_{2}/\sqrt{t}$,
where $t$ indicates exposure time.
In the exposure time range of 50--200~s, 
we took the average of the best-fit noise profiles of the individual epochs.
The average noise level was then scaled down by a factor of $\sqrt{3}$,
to account for the co-adding of the three epochs. Finally, 
this value was used to identify significant clumps in the co-added image.
Though each image has the same exposure time, the data quality also depends on 
the amount of precipitable water vapor in the sky during the observations 
as well as on the elevation of the field. 
As shown in Figure~\ref{fig:exptime}, however, 
the data points over the three epoch are consistent, 
implying that the data quality is comparable from epoch to epoch.
We measured the RMS noise values for each of the 12 co-added images in order to perform 
clump identification.
For the 12 co-added images,
the averaged mean value of the resultant RMS noise levels for finding clumps is
$\sim 4$~\mJyperbm.  
In a single epoch image, the RMS noise level reaches $\sim 8$~\mJyperbm\ 
in the central area with the longest exposure time.

\subsection{Measuring the Flux: Step 2} \label{sec:reduc-s2}

We measured the peak flux $\fe(i)$ for each clump $i$ and epoch $e$. 
We denote the  mean peak flux over the three epochs as $\fmean(i)$. 
The peak flux measurements are robust as the 8\arcsec\ Gaussian smoothing mitigates 
pixel-to-pixel noise variations and 
the peak position uncertainty from epoch to epoch is less than beam size.
In addition, we selected clumps with $\fmean(i) \geq 250$~\mJyperbm\ for this analysis, 
which is $\sim25$~S/N in a single epoch (noise $\sim10$~\mJyperbm).
To find stable calibrator sources for relative flux calibration,
we first assumed that all clumps are not variable.
In each epoch $e$, we derived a RFCF as follows:
\begin{equation}
{\rm RFCF} = \frac{\sum_{c=1}^{\nc}\fe(c)/\fmean(c)}{\nc},
\label{eq:rfcf}
\end{equation}
where $c$ denotes a calibrator, and $\nc$ is the number of calibrators per field. 
Each epoch image was divided by its RFCF in order to calibrate the images relative to one another.
From these relative flux calibrated images,
we remeasured the peak fluxes in each epoch
and compared the standard deviation, $\sd(i)$, of the clump fluxes with a 
fiducial standard deviation model, $\sdfid(i)$.
The fiducial standard deviation model characterizes the uncertainty 
in a relative flux calibrated image based on the RMS noise ($\noise(i)$) 
and the relative flux calibration uncertainty  
itself ($\ucal$; see  \citealt{johnstone2018} for further details). 
$\sdfid(i)$ is calculated as follows:
\begin{equation}
\sdfid(i) = \sqrt{\noise(i)^2+(\ucal \times \fmean(i))^2}, 
\label{eq:sdfid}
\end{equation}
where $\ucal$ is 
\begin{equation}
	\ucal = \sqrt{\frac{\sum_{c=1}^{\nc}\sd(c)^2/\fmean(c)^2}{\nc-1}}.
\label{eq:ucal}
\end{equation}
Here, $\noise(i)$ is the mean value of the three epoch noise levels shown in Table~\ref{tab:out},
and $\ucal$ is given in the last column of Table~\ref{tab:cal}.

\begin{figure*}[ht!]
\epsscale{.8}
\plotone{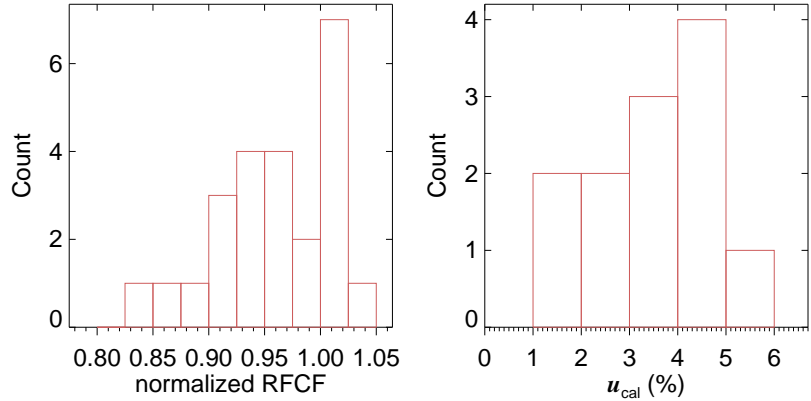}
\caption{
Histograms of the RFCF
and the relative calibration uncertainty, $\ucal$, from Table~\ref{tab:cal}.
Values in the RFCF histogram were normalized 
to the first epoch for each field,
so only RFCFs derived for the other epochs are counted.
\label{fig:rfcf}}
\end{figure*}

The relative calibration steps were repeated using a clipping process 
to identify a set of stable calibrators.
After applying the relative flux conversions for each epoch,
we compared the expected uncertainty for each source ($\sdfid(i)$)
with the measured value ($\sd(i)$).
As discussed in Section~\ref{sec:res-s1}, with only three measurements,
we expected $\sd < 1.7 \times \sdfid$,
which corresponds to a 95\% of confidence level if there is no intrinsic variability.
The numbers of identified clumps, calibrator sources, outliers,
the RFCF at each epoch, and $\ucal$ are listed in Table~\ref{tab:cal}
(see Section~\ref{sec:res-s1} for details on the outliers).
Figure~\ref{fig:rfcf} shows
 histograms of the normalized RFCFs (normalized to the first epoch) 
and associated uncertainties.
The normalized RFCFs were 
used to moderate the effects of small number statistics. 
The applied RFCFs were within the nominal flux calibration uncertainty 
of SCUBA-2 data at 850~\um\ \citep{dempsey2013, mairs2017b}.
The median relative calibration uncertainty ($\ucal$) was found to be $\sim 3.6$\%, 
which is slightly higher than what the Transient Survey team achieved ($\sim 2$\%).
This slight increase in the relative calibration uncertainty is primarily
due to two effects: the lower brightness limit used here for potential calibrators
and the necessity to allow extended sources as calibrators.
These differences from the Transient Survey are discussed in more detail below.

\subsection{Differences in Methodology from the Transient Survey}
 
We have adopted the methods performed by the Transient Survey team
to investigate peak flux changes over time.
However, the SCOPE survey was not optimized for this type of work,
so the following alterations to the Transient Survey methodology were applied.

First, our smoothing kernel size is slightly larger than that of the Transient Survey team (8\arcsec\ as opposed to 6\arcsec).
Second, we used the \clumpfind\ algorithm 
while the Transient Survey team used {\sc{Gaussclumps}} \citep{stutzki1990}. 
Both of these algorithms provide almost the same results overall,
but there are some differences in complex areas of a given map.
Third, we applied a different set of criteria from the Transient Survey to select clumps from the catalogs obtained by using each algorithm.
The Transient Survey team considered 
only sources which are very bright ($> 50\,\noise$) and compact 
(effective radius assuming a circular projected configuration $< 10\arcsec$), 
and which appear in every epoch. 
Alternatively,
we included less bright ($> 25\,\noise$) sources and more extended sources.
Fourth, 
the calibrator selection described above in this section differs from
that of the Transient Survey team 
due to the difference in the number of bright sources.
While we considered all the clumps to be potential calibrators at the beginning 
and then selected the invariable clumps, 
the Transient Survey team could be more selective as their fields contain 
many compact, bright clumps for the calibration such that 
the uncertainty from the noise was less than 5\% \citep{mairs2017b}.  
In spite of the differences in bright source selection for the relative flux calibration,
the procedure presented in this study is sufficient 
to detect a flux variation of 10\% ($3\times\ucal$).

\section{Results} \label{sec:res}

\subsection{Analysis of Peak Flux Measurement} \label{sec:res-s1}

\begin{figure*}[ht!]
\epsscale{1}
\plotone{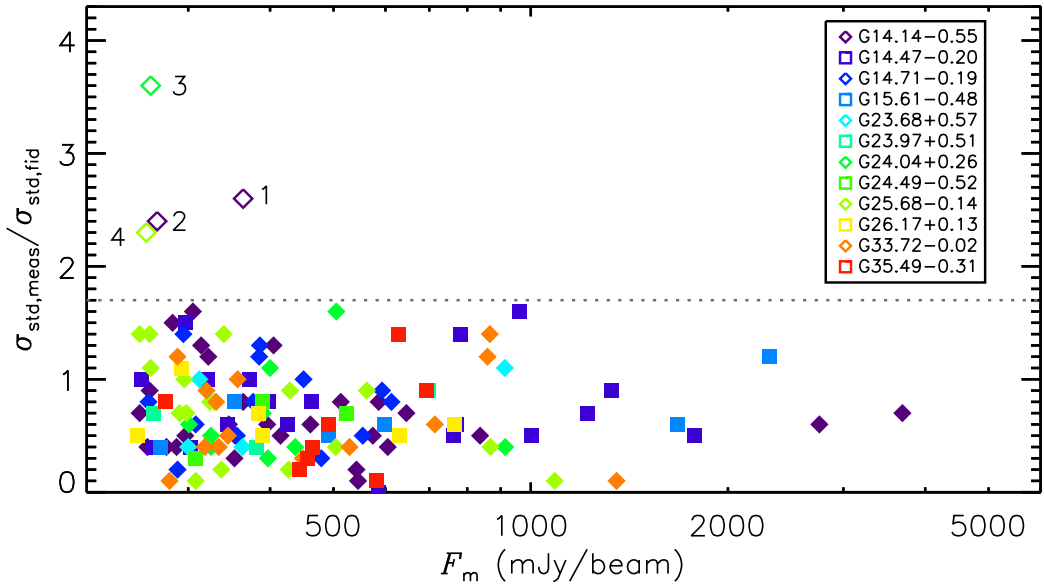}
\caption{
$\sd/\sdfid$ versus $\fmean$ for all identified clumps.
Filled and open symbols are calibrators and outliers, respectively.
A dashed line marks a threshold of 1.7 for calibrator sources (see the text for more details).
The number assigned to each outlier is also marked.
\label{fig:sig}}
\end{figure*}

We identified 136 clumps with $\fmean \geq 250$~\mJyperbm\ across the 12 fields.
Figure~\ref{fig:sig} 
shows the $\sd/\sdfid$ as a function of the mean peak flux density.
Almost all clumps (132/136; marked with filled symbols in the figure) 
show little flux changes and are used as calibrators.
Four outliers (open symbols) in three different SCOPE fields were detected.

\citet{johnstone2018} searched for submillimeter variability in 1643 bright sources 
across eight star-forming regions using the first 18-month data
of monthly observations obtained by the JCMT Transient Survey.
Figure~2 of \citet{johnstone2018} is similar to Figure~\ref{fig:sig} in this paper.
Their results of $\sd/\sdfid$ are much more tightly constrained toward a value of 1.
This is mainly due to their larger set of data (10--15 epochs) per region. 
EC~53, a known variable source in Serpens Main \citep{hodapp2012, yoo2017}, 
is an extreme outlier with a value of $\sd/\sdfid =$ 5.6.
We found no clump that shows similar, exceptional variability in our data.

To analyze how significant the outlier detections are, 
we constructed a simple statistical test of the null hypothesis that
there is no variability beyond the flux changes due to the observational uncertainty.
For 100,000 trials, we drew three peak values (to represent three epochs) at random from a normal distribution
with a mean of a given peak value and 
a standard deviation of $\sdfid$. 
We measured $\sd$ from these three measurements, calculated  
$\sd/\sdfid$ for each trial,
and examined the probability density function of $\sd/\sdfid$.
We found that the probability density function depends only on the number of observational epochs.
For the three epoch case, the $\sd/\sdfid$ distribution has a mean of 0.85 and a median of 0.83.
$\sd/\sdfid \simeq$ 1.7 and 3.1 give
the cumulative probabilities of $\sim 95\%$ and $\gtrsim 99.99\%$, respectively. 
100\% minus the cumulative probability indicates 
the probability that the flux changes are simply due to the observational uncertainty.
All four outliers in Figure~\ref{fig:sig} have $\sd/\sdfid \ge 2.3$,
which corresponds to less than 0.5\%.  
(This result is equivalent to identifying outliers at least $2.8\sigma$ from the mean in a normal distribution.) 
Therefore, they might be candidate variable sources.

The four outliers are listed in Table~\ref{tab:out}.
The parameter of $\sd/\sdfid$ is a good, dimensionless indicator of flux variability.
For the outliers, $\sd/\sdfid$ is between 2.3 and 3.6.
Compared with EC~53,
the outliers have much smaller values of $\sd/\sdfid$.
In addition, all four outliers are relatively faint clumps ($\lesssim 400$~\mJyperbm).
The outliers are described below in more detail.

\begin{figure}[ht!]
\epsscale{.9}
\plotone{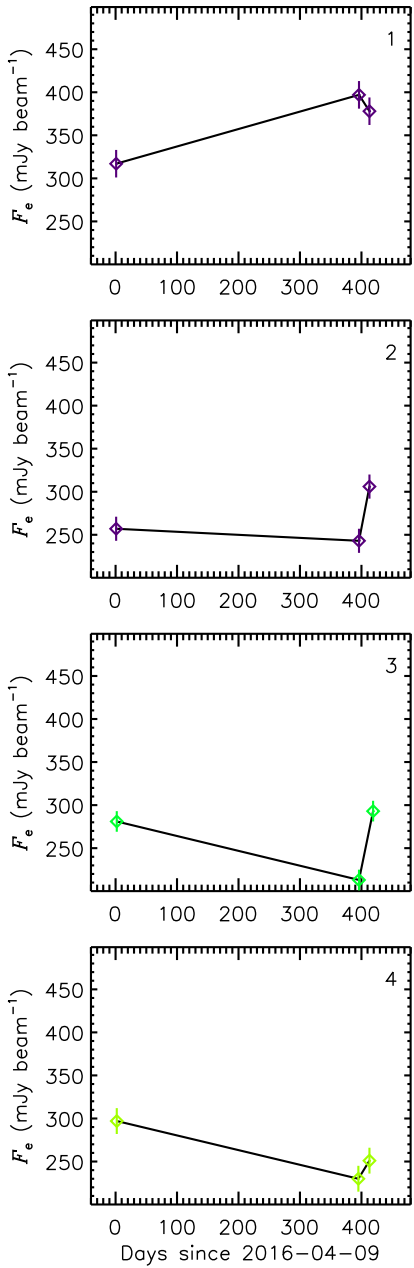}
\caption{Peak flux variations of the four selected outliers in Table~\ref{tab:out}.
The number assigned to each outlier is written in the top right corner.
Symbols and colors are described in Figure~\ref{fig:sig}.
The error bars represent $\sdfid$.
\label{fig:lc}}
\end{figure}

\begin{figure*}
\epsscale{.79}
\plotone{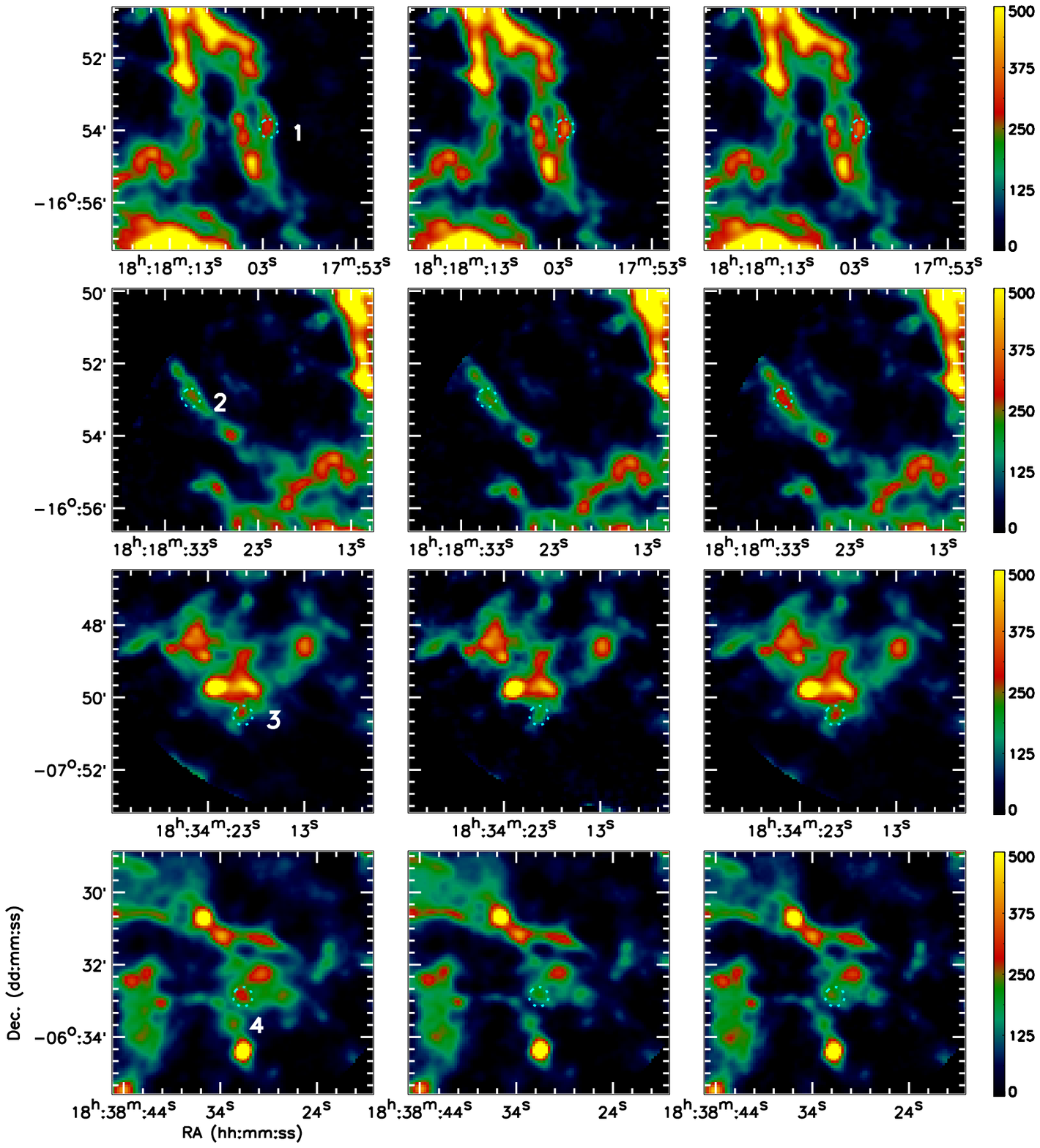}
\caption{
JCMT 850~\um\ images of the four outliers (top to bottom)
listed in Table~\ref{tab:out} for each of the three epochs (left to right).
The images have been re-scaled by the relative calibration process
described in Section~\ref{sec:reduc}.
The outliers are marked using 15\arcsec-radius dotted circles in cyan.
The assigned numbers are written in the right side of the circles in the leftmost images.
The color scale corresponds to the 850~\um\ flux in \mJyperbm.
\label{fig:epoch}}
\end{figure*}

Figures~\ref{fig:lc} and \ref{fig:epoch}
show the peak flux variations of the outliers at 850~\um\ 
which are approximately six times the noise level.
While it is difficult to define the variability timescale
with a limited number of observations and an uneven observational cadence,
we analyzed the trend of peak fluxes.
Outliers 1 and 4 showed clear differences between the first and the two subsequent epochs.
Outlier~2 showed no significant flux variations 
between the first two epochs separated by a year,
but a sudden flux increase is detected between the second and third epochs 
separated by less than a month.
Outlier~3 showed a clear difference in flux after 
the initial long time interval and also after the later, shorter time interval.
The fluxes measured in the first and last epochs, however, were similar to one another.
The JCMT Transient Survey found that
the majority of variables uncovered have long-term (a number of years),
rather than short-term variations (monthly-to-yearly timescales)
\citep{johnstone2018}, 
though only rare, extremely bright events allow the survey to uncover variations
within individual epochs \citep{mairs2019}.
Further monitoring is required to confirm 
such short-term variations.

\subsection{Large-Scale Bias Check} \label{sec:res-s2}

Thus far, the technique we used in this paper is 
to compare the peak fluxes of different epochs for each clump after relative flux calibration.
However, it is well known that submillimeter continuum map reconstruction often creates 
low-level, artificial, extended structures that may affect simple peak flux measurements.
Such complications are more likely to arise across small crowded maps,
such as those undertaken by SCOPE, as compared with the large, sparser Transient Survey fields.
Thus, in this section we test whether the observed brightness variations from the four candidate variables are truly localized as expected for compact sources.

\begin{figure*}
\epsscale{.79}
\plotone{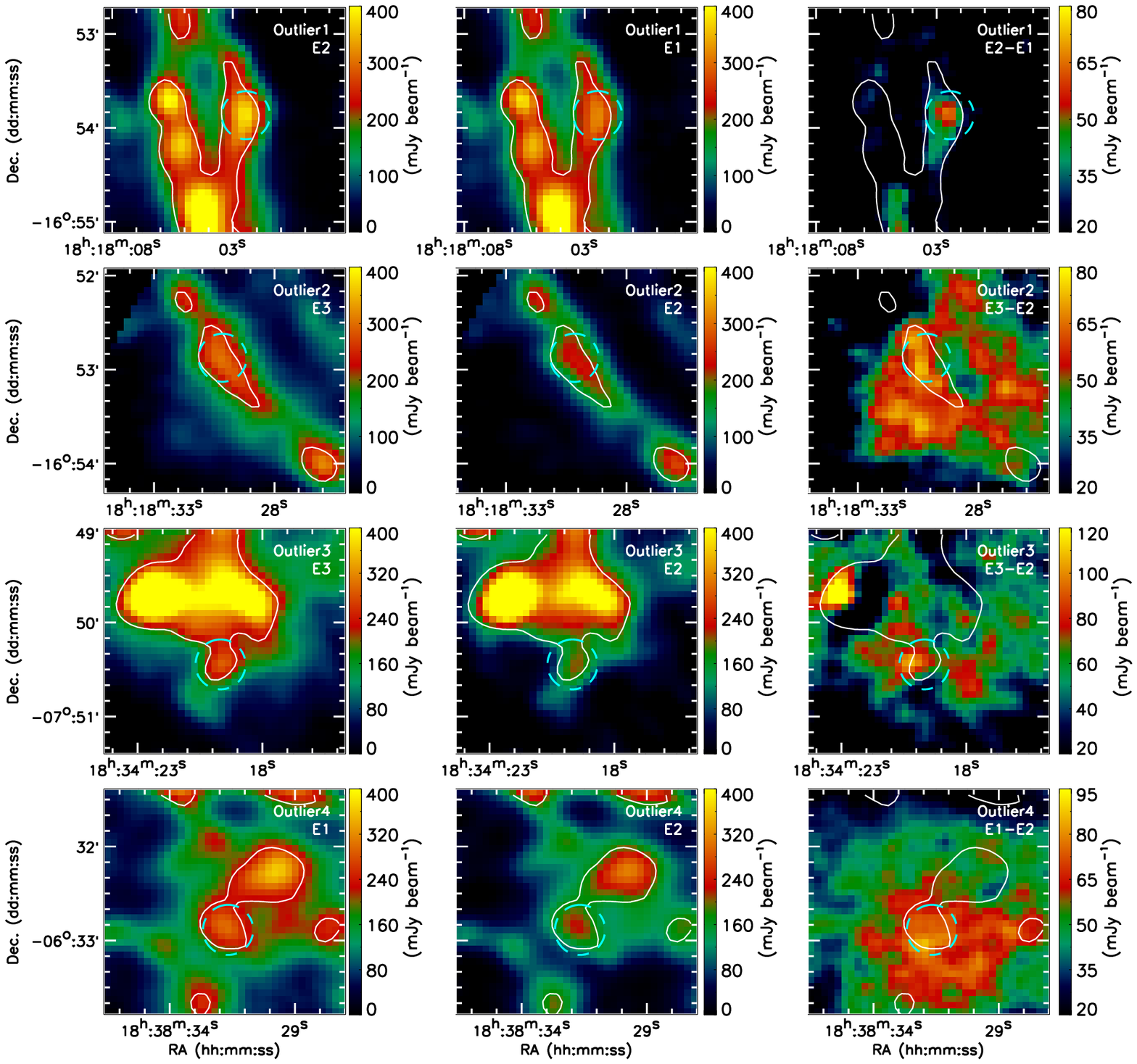}
\caption{
Flux difference maps of the four outliers at JCMT 850~\um.
For each outlier, there are three images: (left) the outlier-brightest epoch image,
(center) the outlier-faintest epoch image, and (right) their difference map.
White contours show a flux level of 230~\mJyperbm\ from the leftmost image. 
The outliers are marked using 15\arcsec-radius dashed circles in cyan.
\label{fig:diff-map}}
\end{figure*}

Thus, we aligned SCOPE images using the algorithm IDL/SUBREG mentioned in Section~\ref{sec:reduc}
and made difference maps using those epochs containing the minimum and maximum peak flux values.
Figure~\ref{fig:diff-map} shows
the flux difference maps of the four outlier candidates,
zoomed in to localized areas of $2\arcmin \times 2\arcmin$.
For each source, there are three panels: brightest and faintest epoch outlier images and their difference map.
For Outlier~1, it appears that the majority of the flux change is located at the peak position.
Therefore, we can confirm that the flux variation genuinely originates from the brightness of the localized source.
On the other hand,
for the other three sources (Outliers~2--4),
between epochs the extended emission rises along with the peak flux increase.
This can be seen most clearly in Outlier~4. 
For Outlier~2, there is a peaking-up trend above the background change by $\sim$30~\mJyperbm,
which is only about half of the anticipated value from the peak flux analysis alone.
For Outlier~3, there is an increase of about 60~\mJyperbm\ over the background change.
However, this trend does not peak at the location of the source.

In summary, we find that three of the four candidate variables (Outliers~2--4) are
closely associated with large-scale flux variations between epochs.
As we do not expect to observe large variations in the brightness of an extended structure in star-forming regions
and we are well aware of the likelihood of artificial large-scale structure
created during the map-making process,
we remove these three sources from any further analysis.
Outlier~1 remains a ``candidate" variable, 
although it is not particularly ``robust" (see Section~\ref{sec:res-s1}).

\section{Discussion} \label{sec:dis}

\subsection{Variable candidate found in this study} \label{sec:outlier}

Outlier~1 (G14.143$-$0.508) was 
found in the G14.14$-$0.55 field. 
The 28 invariable clumps in the field have 
$\sd/\sdfid \simeq 0.1$--1.6 with an average of 0.7, 
while the outlier has 2.6.

We investigated whether Outlier~1 shows signs of star formation,
in which case the detected flux change could potentially be attributed to accretion variability.
The clumps we identify in this study were covered by
the APEX Telescope Large Area Survey of the Galaxy (ATLASGAL; \citealt{schuller2009}). 
We, therefore, searched for ATLASGAL clumps near the peak flux position of this outlier. 
\citet{urquhart2018} derived the distances and physical properties  
(including evolutionary classification) 
of about 8000 ATLASGAL clumps in Galactic disk in the Galactic longitude from 5\arcdeg\ to 60\arcdeg.
The ATLASGAL was conducted at 870~\um\ with a beam size of 19\farcs2. 
The observing wavelength and beam size are comparable to ours.
Note that the ATLASGAL survey has a typical noise level of 50--70~\mJyperbm,
which is one order of magnitude higher than that of the SCOPE survey.
Outlier~1 is associated with ATLASGAL clump AGAL014.142$-$00.509
that has a $\vlsr$ of 21.1~\kms.
The kinematic distance was estimated to be 1.5~kpc \citep{urquhart2018}.

This clump seems to be deeply embedded in an IR dark cloud filament.
\citet{urquhart2018} inferred Outlier~1 to be in a quiescent phase,
because it is dark or weak at near- to far-IR wavelengths.
The flux variation in a quiescent (seemingly starless) clump clump may sound contradictory. 
It can be explained, however, by
the presence of at least one undetected heavily embedded (proto)star(s).
For example,
recent studies by \citet{liu2018b} using single-dish telescopes and \citet{contreras2018} 
using the Atacama Large Millimeter/submillimeter Array (ALMA) detected 
high accretion rates in massive quiescent cores,
which are comparable to those found in high-mass protostellar objects \citep[see also][]{traficante2017}.
Also, the non-detection of an IR counterpart
may be due to the sensitivity limits of existing mid/far-IR surveys
\citep[e.g., see Section 3.2.4 of][]{svoboda2016}.
\citet{urquhart2019} found from a molecular line survey that $\sim 35\%$ of 29 quiescent ATLASGAL clumps
in their sample have relatively high (30--50~K) rotation temperatures, 
suggesting the existence of internal heating protostellar object(s).
Moreover, there are indeed several discoveries of compact bipolar molecular outflows
in otherwise quiescent ATLASGAL clumps/cores \citep[e.g.,][]{feng2016, tan2016, pillai2019}.
The driving sources of the detected outflows were suggested to be massive protostars in the very early evolutionary stage.
If Outlier~1 is a seemingly starless clump (in fact, not starless),
high-resolution molecular observations could uncover that the clump is actually in the earliest stage of star formation.
Therefore, a further investigation of this clump
at higher angular resolution and sensitivity is required to uncover the embedded protostar(s).

The relationship between the bolometric luminosity and the envelope mass
is useful for determining whether there is low- or high-mass star formation occurring
\citep[e.g.,][]{molinari2008,urquhart2014,motte2018}.
Based on the luminosity ($\simeq 37~\Lsun$) and the mass ($\simeq 23~\Msun$) of the associated clump,
Outlier~1 is very likely related to low- or intermediate-mass star formation 
rather than high-mass star formation.

\subsection{Comparison with Known Submillimeter Variable Sources}

There are a few known submillimeter variable sources 
observed in low- and high-mass star-forming regions.
They are relatively close ($\lesssim 1.3$~kpc),
while Outlier~1 in this study appears to be slightly more distant.
As an example of a low-mass, variable protostellar system,
EC~53 is located at 436~pc \citep{ortiz2017},
and it brightened from 960 to 1450~\mJyperbm\ at 850~\um\ 
in a 14\farcs6 single-dish beam \citep{yoo2017}.
If moved to a greater distance, the brightness of EC~53 will diminish 
significantly as the distance increases and, thus, it would fall below our sensitivity threshold.
The area of an outburst associated with accretion variability is 
unresolved at the outliers' distances.
Even if it were embedded in additional material,
allowing the larger clump to be visible, 
the change in brightness of EC~53 would only be
$\sim 40$ or 1~mJy at far distances of 1.5 or 10~kpc, respectively.
These low flux variations are marginally detectable or undetectable levels in our observations.
Alternatively,
the massive protostellar system NGC~6334I-MM1 at 1.3~kpc \citep{chibueze2014, reid2014} 
brightened by as much as 30\% of the single-dish flux
($= 0.3 \times 65300$ $\simeq 19600$~\mJyperbm)
at 850~\um\ in an 17\farcs5 beam
\citep{sandell1994,hunter2017}.
In the same way, 
one expects to measure the increase of $\sim 14700$ and 300~mJy
for an event like NGC~6334I-MM1 at distances of 1.5 and 10~kpc, respectively.
These large variations would be easily detectable in our observations,
but none of our SCOPE sources show such a dramatic change.
The peak flux change of Outlier~1, thus, appears to be 
related to an event of an intermediate scale in terms of luminosity 
based on the EC~53 and NGC~3664I-MM1 case analyses.

This study suggests that
long-term monitoring of distant star-forming regions
with the JCMT is suitable for detecting submillimeter variability.
If the variable candidate is confirmed,
observations with higher angular resolution and sensitivity, using an interferometer such as the ALMA,
will give us a better understanding of their properties.
Follow-up high-resolution observations will not only 
more easily detect any flux variability with little beam dilution
but they will also reveal the embedded young stellar object(s) being responsible for variability events.
As an example, 
the interferometric flux of NGC~6334I-MM1 increased by a factor of 4,
which is much greater than a $\sim 30\%$ increase in the single-dish flux \citep{hunter2017}.

\section{Summary and Conclusion} \label{sec:sum}

We investigated the flux variations of submillimeter clumps in
12 PGCC fields in the first quadrant of the Galactic plane.
The fields were observed three times over approximately 14 months
using the JCMT/SCUBA-2, as part of the SCOPE survey.
The survey was not optimized for detailed studies on flux variation
and, therefore, the observations only cover three epochs
with uneven time intervals.
Nevertheless, taking into account the non-uniform noise distributions of the maps,
we succeeded in examining relative flux changes 
by comparing the peak fluxes of identified clumps among epochs. 
We performed a relative flux calibration as described in Section~\ref{sec:reduc},
with a typical uncertainty of $\sim 3.6\%$.
In the 12 PGCC fields,
we identified 136 clumps with mean peak flux densities larger than 250~\mJyperbm\ 
($\gtrsim$ 25 S/N).
From the peak flux analysis,
we found four ``outliers" that appear to vary in time.
The average flux change at 850~\um\ is about 30\%.
We examined whether the peak flux changes of the outliers are well localized in the flux difference maps.
Finally, only one (Outlier~1) of the four outliers is a plausible ``candidate" and is not biased by the large scales.
The detected flux variation in Outlier~1 may be related to episodic accretion events
in the very early stage of low- or intermediate-mass star formation,
considering a kinematic distance of 1.5~kpc,
although we cannot completely exclude the possibility that it is a purely statistical random deviation.
According to the existing observational data at near- to far-IR,
the star-forming sign is less evident.
However, the flux variability found here suggests
an additional investigation of this region at higher angular resolution
and sensitivity to uncover the deeply embedded protostar(s) in this clump.
Further research employing long-term monitoring will be helpful 
not only to confirm our results 
but also to give a better understanding of the accretion processes in star formation.


\input{tab1.tex} 
\input{tab2.tex}

\input{tab3.tex}

\acknowledgments

We thank the anonymous referee for many valuable comments and suggestions which helped us improve the paper.
D.J. is supported by NRC Canada and an NSERC Discovery Grant. 
T.L. is supported by the EACOA fellowship.
J.E.L. is supported by the Basic Science Research Program through the National Research Foundation of Korea (grant No. NRF-2018R1A2B6003423) and the Korea Astronomy and Space Science Institute under the R\&D program supervised by the Ministry of Science, ICT and Future Planning.
M.J. acknowledges the support of the Academy of Finland Grant No. 285769.
M.K. is supported by Basic Science Research Program through the National Research Foundation of Korea (NRF) funded by the Ministry of Science, ICT, and Future Planning (No. NRF-2015R1C1A1A01052160).
J.M. acknowledges the support by the Programme National ``Physique et Chimie du Milieu Interstellaire" (PCMI) of CNRS/INSU with INC/INP co-funded by CEA and CNES.
A.S. is supported by National Science Foundation Grant-1715876.
C.W.L. is supported by the Basic Science Research Program through the National Research Foundation of Korea (NRF) 
funded by the Ministry of Education, Science and Technology (NRF-2019R1A2C1010851).
W.K. is supported by Basic Science Research Program through the National Research Foundation of Korea (NRF-2016R1C1B2013642).
The James Clerk Maxwell Telescope is operated by the East Asian Observatory 
on behalf of The National Astronomical Observatory of Japan; 
Academia Sinica Institute of Astronomy and Astrophysics; 
the Korea Astronomy and Space Science Institute; 
the Operation, Maintenance and Upgrading Fund 
for Astronomical Telescopes and Facility Instruments,
budgeted from the Ministry of Finance (MOF) of China 
and administrated by the Chinese Academy of Sciences (CAS),
as well as the National Key R\&D Program of China (No. 2017YFA0402700). 
Additional funding support is provided by the Science and Technology Facilities Council 
of the United Kingdom and participating universities in the United Kingdom and Canada.
The identification number for the JCMT SCOPE Survey 
under which the SCUBA-2 data were used in this paper is M16AL003.



\vspace{5mm}
\facilities{JCMT (SCUBA-2) \citep{holland2013}}

\software{\starlink\ \citep{currie2014}, \idl}

\end{document}

%% file: tab1.tex
\begin{deluxetable*}{crrcc cccc}
\tabletypesize{\scriptsize}
\tablecaption{Fields and Epochs \label{tab:fieldinfo}}
\tablehead{
\colhead{} & \multicolumn{2}{c}{Central Position\tablenotemark{a}} && \colhead{Three Epochs} &&
\multicolumn{2}{c}{Time Intervals\tablenotemark{b}} & \colhead{Distance(s)\tablenotemark{c}} \\
\cline{2-3} \cline{5-5} \cline{7-8}
\colhead{Field} & \colhead{(h:m:s)} & \colhead{(d:m:s)} && \colhead{(yyyymmdd)} && 
\multicolumn{2}{c}{(day)} & \colhead{(kpc)}
}
\startdata
 G14.14$-$0.55 &18:18:11.50& $-$16:55:29.05 && 20160410~~20170510~~20170527 && 395 & ~17 &  1.5 \\ 
 G14.47$-$0.20 &18:17:31.80& $-$16:28:00.46 && 20160409~~20170511~~20170602 && 397 & ~22 &  3.1 (11.5) \\
 G14.71$-$0.19 &18:17:59.80& $-$16:14:41.16 && 20160409~~20170510~~20170602 && 396 & ~23 &  3.1 \\ 
 G15.61$-$0.48 &18:20:48.40& $-$15:35:41.29 && 20160410~~20170511~~20170602 && 396 & ~22 &  1.8 and 16.9 \\ 
 G23.68$+$0.57 &18:32:23.20& $-$07:57:39.50 && 20160411~~20170510~~20170603 && 394 & ~24 &  5.8 \\ 
 G23.97$+$0.51 &18:33:09.20& $-$07:43:48.16 && 20160411~~20170512~~20170604 && 396 & ~23 &  5.8 \\ 
 G24.04$+$0.26 &18:34:10.40& $-$07:47:05.86 && 20160411~~20170510~~20170602 && 394 & ~23 &  7.8 \\ 
 G24.49$-$0.52 &18:37:48.10& $-$07:44:45.61 && 20160411~~20170512~~20170602 && 396 & ~21 & 11.3 \\ 
 G25.68$-$0.14 &18:38:39.10& $-$06:30:49.20 && 20160411~~20170509~~20170527 && 393 & ~18 & 10.2 (7.4) \\ 
 G26.17$+$0.13 &18:38:34.70& $-$05:57:20.53 && 20160411~~20160830~~20170604 && 141 & 278 &  7.6 \\ 
 G33.72$-$0.02 &18:52:55.20& $+$00:41:26.00 && 20160412~~20160722~~20170527 && 101 & 309 &  6.5 (2.2) \\ 
 G35.49$-$0.31 &18:57:12.90& $+$02:07:52.72 && 20160413~~20160607~~20170527 && ~55 & 354 &  2.7 (3.2 and 10.3) \\ 
\enddata                                                                                          
\tablenotetext{a}{Equatorial coordinates, R.A. and decl. (J2000)
	}
\tablenotetext{b}{Time intervals between the first and second epochs and between the second and third epochs.
	}
\tablenotetext{c}{Distances are obtained from \citet[][see also references therein]{urquhart2018}. 
	For fields having clumps at various distances,
	we give the distance of the majority of clumps along with the value(s) of the minority in parenthesis,
	or, if they are almost equal numbers, two values with the conjunction ``and."
	}
\end{deluxetable*}


%% file: tab2.tex
\begin{deluxetable*}{ccccc ccc}
\tabletypesize{\scriptsize}
\tablecaption{Number of Clumps Found and Relative Calibration Information\label{tab:cal}}
\tablehead{
\colhead{} & \colhead{All Clumps Found} &\colhead{}& \colhead{}&
\multicolumn{3}{c}{RFCF at Each Epoch} & \colhead{$\ucal$} \\
\cline{5-7}
\colhead{Field} & \colhead{$> 250$~\mJyperbm} & \colhead{Calibrators} & \colhead{Outliers} &
\colhead{First} & \colhead{Second} & \colhead{Third} &  \colhead{(\%)}
}
\startdata
G14.14$-$0.55 & 30 & 28 & 2  & 1.005 & 0.981 & 1.014 & 3.6 \\  
G14.47$-$0.20 & 19 & 19 & 0  & 1.014 & 0.953 & 1.033 & 4.5 \\ 
G14.71$-$0.19 & 13 & 13 & 0  & 1.029 & 0.932 & 1.039 & 5.2 \\ 
G15.61$-$0.48 &  6 &  6 & 0  & 0.994 & 0.995 & 1.010 & 1.8 \\       
G23.68$+$0.57 &  4 &  4 & 0  & 1.027 & 0.937 & 1.037 & 4.2 \\ 
G23.97$+$0.51 &  3 &  3 & 0  & 1.005 & 0.988 & 1.010 & 2.8 \\  
G24.04$+$0.26 & 10 &  9 & 1  & 1.038 & 0.895 & 1.066 & 3.1 \\        
G24.49$-$0.52 &  4 &  4 & 0  & 1.025 & 0.981 & 0.995 & 4.9 \\ 
G25.68$-$0.14 & 18 & 17 & 1  & 1.043 & 1.012 & 0.945 & 4.4 \\
G26.17$+$0.13 &  6 &  6 & 0  & 1.085 & 0.902 & 1.013 & 3.4 \\ 
G33.72$-$0.02 & 14 & 14 & 0  & 1.033 & 0.992 & 0.975 & 2.5 \\ 
G35.49$-$0.31 &  9 &  9 & 0  & 1.056 & 0.948 & 0.996 & 1.7 \\ 
\enddata                   
\end{deluxetable*}

%% file: tab3.tex
\begin{deluxetable*}{lcc ccc rcrcrccc cccc}
\tabletypesize{\scriptsize}
\setlength{\tabcolsep}{0.04in}
\tablecaption{Peak flux of Outliers in 850~\um\ \label{tab:out}}
\tablewidth{0pt}
\tablehead{
\colhead{} & \colhead{} &\colhead{} & \multicolumn{2}{c}{Peak Position\tablenotemark{a}} &&  
\multicolumn{6}{c}{$\fe$ at Each Epoch\tablenotemark{b,c}} && 
\colhead{} & \colhead{} &
\colhead{} & \colhead{}\\
\cline{4-5} \cline{7-12}
\colhead{\#} & \colhead{Field} & \colhead{Name\tablenotemark{a}} & \colhead{R.A.(J2000)} & \colhead{Decl. (J2000)} &&
\multicolumn{2}{c}{First} & \multicolumn{2}{c}{Second} & \multicolumn{2}{c}{Third} && 
\colhead{$\fmean$\tablenotemark{c}} & \colhead{$\sd$\tablenotemark{c}} & 
\colhead{$\sdfid$\tablenotemark{c,d}} & \colhead{$\frac{\sd}{\sdfid}$\tablenotemark{e}}
}
\startdata
 1& G14.14$-$0.55 &G14.143$-$0.508 & 18:18:02.02 &$-$16:53:57.09  &&  317&(10)& 397&(10)& 378&(10)&& 364 & 42 & 16 & 2.6 ($\sim 0.1$\%)  \\ 
 2& G14.14$-$0.55 &G14.210$-$0.598 & 18:18:29.89 &$-$16:52:57.05  &&  257&(11)& 243&(10)& 306&(10)&& 269 & 33 & 14 & 2.4 ($\sim 0.3$\%) \\ 
 3& G24.04$+$0.26 &G24.008$+$0.203 & 18:34:19.82 &$-$07:50:29.89  &&  281&(8) & 213&(11)& 293&(8) && 263 & 43 & 12 & 3.6 ($< 0.01$\%) \\ 
 4& G25.68$-$0.14 &G25.635$-$0.126 & 18:38:31.32 &$-$06:32:53.20  &&  297&(9) & 230&(8) & 251&(11)&& 259 & 34 & 15 & 2.3 ($\sim 0.5$\%) \\ 
\enddata                   
\tablenotetext{a}{Name contains each peak position in Galactic coordinates.
	It is determined from the epoch data with the highest peak flux.}
\tablenotetext{b}{Values in parentheses are map noise levels.}
\tablenotetext{c}{Units of \mJyperbm.}
\tablenotetext{d}{see Equation~(\ref{eq:sdfid}). 
	For each source, the mean of noise levels at three epochs and
	$\ucal$ listed in Table~\ref{tab:cal} are used.}
\tablenotetext{e}{Values in parentheses indicate how reliable
    the explanation that the flux change is due to the observational uncertainty is
    (See Section~\ref{sec:res}).}
\end{deluxetable*}

%% file: main.bbl
\begin{thebibliography}{}
\bibitem[Aspin et al.(2009)]{aspin2009} Aspin, C., Reipurth, B., Beck, T.~L., et al.\ 2009, \apjl, 692, L67 
\bibitem[Aspin et al.(2010)]{aspin2010} Aspin, C., Reipurth, B., Herczeg, G.~J., \& Capak, P.\ 2010, \apjl, 719, L50
\bibitem[Bae et al.(2014)]{bae2014} Bae, J., Hartmann, L., Zhu, Z., \& Nelson, R.~P.\ 2014, \apj, 795, 61 
\bibitem[Berry et al.(2007)]{berry2007} Berry, D.~S., Reinhold, K., Jenness, T., \& Economou, F.\ 2007, Astronomical Data Analysis Software and Systems XVI, 376, 425
\bibitem[Caratti o Garatti et al.(2011)]{caratti2011} Caratti o Garatti, A., Garcia Lopez, R., Scholz, A., et al.\ 2011, \aap, 526, L1
\bibitem[Chapin et al.(2013)]{chapin2013} Chapin, E.~L., Berry, D.~S., Gibb, A.~G., et al.\ 2013, \mnras, 430, 2545 
\bibitem[Chibueze et al.(2014)]{chibueze2014} Chibueze, J.~O., Omodaka, T., Handa, T., et al.\ 2014, \apj, 784, 114
\bibitem[Contreras et al.(2018)]{contreras2018} Contreras, Y., Sanhueza, P., Jackson, J.~M., et al.\ 2018, \apj, 861, 14 
\bibitem[Covey et al.(2011)]{covey2011} Covey, K.~R., Hillenbrand, L.~A., Miller, A.~A., et al.\ 2011, \aj, 141, 40
\bibitem[Currie et al.(2014)]{currie2014} Currie, M.~J., Berry, D.~S., Jenness, T., et al.\ 2014, Astronomical Data Analysis Software and Systems XXIII, 485, 391
\bibitem[Dempsey et al.(2013)]{dempsey2013} Dempsey, J.~T., Friberg, P., Jenness, T., et al.\ 2013, \mnras, 430, 2534 
\bibitem[Dunham et al.(2010)]{dunham2010} Dunham, M.~M., Evans, N.~J., II, Terebey, S., Dullemond, C.~P., \& Young, C.~H.\ 2010, \apj, 710, 470-502 
\bibitem[Dunham \& Vorobyov(2012)]{dunham2012} Dunham, M.~M., \& Vorobyov, E.~I.\ 2012, \apj, 747, 52
\bibitem[Eden et al.(2019)]{eden2019} Eden, D., Liu, Tie, Kim, K.-T., et al. 2019, MNRAS, 485, 2895
\bibitem[Evans et al.(2009)]{evans2009} Evans, N.~J., II, Dunham, M.~M., J{\o}rgensen, J.~K., et al.\ 2009, \apjs, 181, 321-350
\bibitem[Fedele et al.(2007)]{fedele2007} Fedele, D., van den Ancker, M.~E., Petr-Gotzens, M.~G., \& Rafanelli, P.\ 2007, \aap, 472, 207
\bibitem[Feng et al.(2016)]{feng2016} Feng, S., Beuther, H., Zhang, Q., et al.\ 2016, \apj, 828, 100
\bibitem[Fischer et al.(2012)]{fischer2012} Fischer, W.~J., Megeath, S.~T., Tobin, J.~J., et al.\ 2012, \apj, 756, 99 
\bibitem[Gibb et al.(2013)]{gibb2013} Gibb, A.~G., Jenness, T., \& Economou, F.\ 2013, PICARD: A PIpeline for Combining and Analyzing Reduced Data, Starlink User Note 265 (Hilo, HI: Joint Astronomy Centre)
\bibitem[Gutermuth et al.(2009)]{gutermuth2009} Gutermuth, R.~A., Megeath, S.~T., Myers, P.~C., et al.\ 2009, \apjs, 184, 18 
\bibitem[Hartmann \& Kenyon(1996)]{hartmann1996} Hartmann, L., \& Kenyon, S.~J.\ 1996, \araa, 34, 207 
\bibitem[Hartmann et al.(2016)]{hartmann2016} Hartmann, L., Herczeg, G., \& Calvet, N.\ 2016, \araa, 54, 135
\bibitem[Herbig(1977)]{herbig1977} Herbig, G.~H.\ 1977, \apj, 217, 693
\bibitem[Herbig(2008)]{herbig2008} Herbig, G.~H.\ 2008, \aj, 135, 637
\bibitem[Herczeg et al.(2017)]{herczeg2017} Herczeg, G.~J., Johnstone, D., Mairs, S., et al.\ 2017, \apj, 849, 43
\bibitem[Hodapp et al.(2012)]{hodapp2012} Hodapp, K.~W., Chini, R., Watermann, R., \& Lemke, R.\ 2012, \apj, 744, 56 
\bibitem[Holland et al.(2013)]{holland2013} Holland, W.~S., Bintley, D., Chapin, E.~L., et al.\ 2013, \mnras, 430, 2513
\bibitem[Hunter et al.(2017)]{hunter2017} Hunter, T.~R., Brogan, C.~L., MacLeod, G., et al.\ 2017, \apjl, 837, L29
\bibitem[Jackson et al.(2006)]{jackson2006} Jackson, J.~M., Rathborne, J.~M., Shah, R.~Y., et al.\ 2006, \apjs, 163, 145
\bibitem[Johnstone et al.(2013)]{johnstone2013} Johnstone, D., Hendricks, B., Herczeg, G.~J., \& Bruderer, S.\ 2013, \apj, 765, 133 
\bibitem[Johnstone et al.(2018)]{johnstone2018} Johnstone, D., Herczeg, G.~J., Mairs, S., et al.\ 2018, \apj, 854, 31 
\bibitem[Liu et al.(2018a)]{liu2018a} Liu, T., Kim, K.-T., Juvela, M., et al.\ 2018a, \apjs, 234, 28
\bibitem[Liu et al.(2018b)]{liu2018b} Liu, T., Li, P.~S., Juvela, M., et al.\ 2018b, \apj, 859, 151 
\bibitem[Kenyon et al.(1990)]{kenyon1990} Kenyon, S.~J., Hartmann, L.~W., Strom, K.~M., \& Strom, S.~E.\ 1990, \aj, 99, 869
\bibitem[K{\'o}sp{\'a}l et al.(2007)]{kospal2007} K{\'o}sp{\'a}l, {\'A}., {\'A}brah{\'a}m, P., Prusti, T., et al.\ 2007, \aap, 470, 211
\bibitem[Mairs et al.(2018)]{mairs2018} Mairs, S., Bell, G.~S., Johnstone, D., et al.\ 2018, The Astronomer's Telegram, 11583
\bibitem[Mairs et al.(2015)]{mairs2015} Mairs, S., Johnstone, D., Kirk, H., et al.\ 2015, \mnras, 454, 2557
\bibitem[Mairs et al.(2017a)]{mairs2017a} Mairs, S., Johnstone, D., Kirk, H., et al.\ 2017a, \apj, 849, 107
\bibitem[Mairs et al.(2019)]{mairs2019} Mairs, S., Lalchand, B., Bower, G.~C., et al.\ 2019, \apj, 871, 72
\bibitem[Mairs et al.(2017b)]{mairs2017b} Mairs, S., Lane, J., Johnstone, D., et al.\ 2017b, \apj, 843, 55
\bibitem[Marton et al.(2016)]{marton2016} Marton, G., T{\'o}th, L.~V., Paladini, R., et al.\ 2016, \mnras, 458, 3479
\bibitem[McKee \& Offner(2011)]{mckee2011} McKee, C.~F., \& Offner, S.~R.~R.\ 2011, , in IAU Symp. 270, Computational Star Formation, ed. J. Alves et al. (Cambridge: Cambridge Univ. Press), 73 
\bibitem[Molinari et al.(2008)]{molinari2008} Molinari, S., Pezzuto, S., Cesaroni, R., et al.\ 2008, \aap, 481, 345.
\bibitem[Motte et al.(2018)]{motte2018} Motte, F., Bontemps, S., \& Louvet, F.\ 2018, \araa, 56, 41
\bibitem[Ortiz-Le{\'o}n et al.(2017)]{ortiz2017} Ortiz-Le{\'o}n, G.~N., Dzib, S.~A., Kounkel, M.~A., et al.\ 2017, \apj, 834, 143
\bibitem[Pillai et al.(2019)]{pillai2019} Pillai, T., Kauffmann, J., Zhang, Q., et al.\ 2019, \aap, 622, A54
\bibitem[Purcell et al.(2012)]{purcell2012} Purcell, C.~R., Longmore, S.~N., Walsh, A.~J., et al.\ 2012, \mnras, 426, 1972 
\bibitem[Planck Collaboration et al.(2016)]{planck2016} Planck Collaboration, Ade, P.~A.~R., Aghanim, N., et al.\ 2016, \aap, 594, A28 
\bibitem[Reid et al.(2014)]{reid2014} Reid, M.~J., Menten, K.~M., Brunthaler, A., et al.\ 2014, \apj, 783, 130
\bibitem[Reipurth et al.(2012)]{reipurth2012} Reipurth, B., Aspin, C., \& Herbig, G.~H.\ 2012, \apjl, 748, L5
\bibitem[Safron et al.(2015)]{safron2015} Safron, E.~J., Fischer, W.~J., Megeath, S.~T., et al.\ 2015, \apjl, 800, L5 
\bibitem[Sandell(1994)]{sandell1994} Sandell, G.\ 1994, \mnras, 271, 75
\bibitem[Schmeja, \& Klessen(2004)]{schmeja2004} Schmeja, S., \& Klessen, R.~S.\ 2004, \aap, 419, 405.
\bibitem[Schuller et al.(2009)]{schuller2009} Schuller, F., Menten, K.~M., Contreras, Y., et al.\ 2009, \aap, 504, 415
\bibitem[Shu(1977)]{shu1977} Shu, F.~H.\ 1977, \apj, 214, 488 
\bibitem[Stutzki \& Guesten(1990)]{stutzki1990} Stutzki, J., \& Guesten, R.\ 1990, \apj, 356, 513
\bibitem[Svoboda et al.(2016)]{svoboda2016} Svoboda, B.~E., Shirley, Y.~L., Battersby, C., et al.\ 2016, \apj, 822, 59
\bibitem[Tan et al.(2016)]{tan2016} Tan, J.~C., Kong, S., Zhang, Y., et al.\ 2016, \apjl, 821, L3
\bibitem[Terebey et al.(1984)]{terebey1984} Terebey, S., Shu, F.~H., \& Cassen, P.\ 1984, \apj, 286, 529
\bibitem[Traficante et al.(2017)]{traficante2017} Traficante, A., Fuller, G.~A., Billot, N., et al.\ 2017, \mnras, 470, 3882
\bibitem[Urquhart et al.(2019)]{urquhart2019} Urquhart, J.~S., Figura, C., Wyrowski, F., et al.\ 2019, \mnras, 484, 4444
\bibitem[Urquhart et al.(2018)]{urquhart2018} Urquhart, J.~S., K{\"o}nig, C., Giannetti, A., et al.\ 2018, \mnras, 473, 1059 
\bibitem[Urquhart et al.(2014)]{urquhart2014} Urquhart, J.~S., Moore, T.~J.~T., Csengeri, T., et al.\ 2014, \mnras, 443, 1555
\bibitem[Vorobyov \& Basu(2015)]{vorobyov2015} Vorobyov, E.~I., \& Basu, S.\ 2015, \apj, 805, 115 
\bibitem[Wienen et al.(2012)]{wienen2012} Wienen, M., Wyrowski, F., Schuller, F., et al.\ 2012, \aap, 544, A146 
\bibitem[Williams et al.(1994)]{williams1994} Williams, J.~P., de Geus, E.~J., \& Blitz, L.\ 1994, \apj, 428, 693 
\bibitem[Whitworth \& Ward-Thompson(2001)]{whitworth2001} Whitworth, A.~P., \& Ward-Thompson, D.\ 2001, \apj, 547, 317
\bibitem[Yoo et al.(2017)]{yoo2017} Yoo, H., Lee, J.-E., Mairs, S., et al.\ 2017, \apj, 849, 69
\end{thebibliography}
